\documentstyle[prb,aps,preprint,tighten]{revtex}
\begin{document}
\title{ Critical properties of gapped spin-$ 1 \over 2$ chains and
ladders in a magnetic field}

\author{R. Chitra and T. Giamarchi}
\address{Laboratoire de Physique des Solides, Universit{\'e} Paris--Sud,
                   B{\^a}t. 510, 91405 Orsay, France}
\maketitle

\begin{abstract}
 An interesting feature of  spin-$1 \over 2$ chains with a gap is that they undergo a
commensurate-incommensurate transition in the presence of an external
magnetic field $H$. The system is in a gapless incommensurate regime for all values of the magnetic field between the  lower critical field $H_{c1}$ and an upper critical field
$H_{c2}$, where it is gapless and has power law correlations. 
 We calculate the critical exponents for such a generic
gapped  system in the incommensurate regime 
at the critical field $H_{c1}$ and in its vicinity. Our analysis also applies to the
spin-$1 \over 2$ ladder. We 
compute the full dynamical susceptibilities at finite temperature.
We use the same to  discuss the thermal broadening of various modes  and 
obtain the low temperature behaviour of the
nuclear spin relaxation rate $T_1^{-1}$. We discuss the
results obtained here  for the special cases of the dimerised chain, frustrated chain
and the spin-$1 \over 2$ ladder.

\end{abstract}

\section{Introduction} 

    Quantum spin chains have been a field of intense theoretical 
activity for the
past decade \cite{AFF}. This seems pertinent in the  face of the tremendous advance
made in the fabrication of quasi-one dimensional systems. 
There are a lot
of compounds  like $ (TMTTF)_2 X$  which are essentially  spin-$1 \over 2$ chains,
NENP - a spin-$1$ chain \cite{RE} and a host of other organic compounds 
whose behaviours can be adequately described within the 
framework of interacting spin systems \cite{S12}.  
The Heisenberg model with purely nearest neighbour interactions has
been instrumental in understanding various properties of  these spin systems \cite{BF}. In addition to these,
there are compounds  with  alternation in the nearest neighbour bonds
 (arises as a result of
interactions with phonons) eg TMTTF, strong single ion anisotropies
etc.
The effects of spin-Peierls
dimerization, interchain interactions
 and competing nearest neighbour interactions have been
reasonably well understood in the spin-$1\over 2$ case \cite{FDM}.
These are known to drive  phase transitions 
in the pure spin-$1 \over 2$ model.
The basic issue addressed is 
 whether there exists  a gap above the singlet ground state
to the first excited state which is a triplet. Apart from these,
the ground state could also exhibit crossovers from
 short range  Neel order to spiral order or spontaneously dimerise \cite{MG}
but these  are not signals of any kind of
quantum phase transitions.

More interesting phases can be obtained by placing these 
 gapped systems in a magnetic field.
For some critical value of the field $H_{c1}$,
the system undergoes a continuous phase transition
 from a commensurate Neel  (C) zero uniform magnetisation
phase to an incommensurate phase (IC) with non-zero magnetisation.
As the magnetic field is increased, the magnetisation increases and
saturates at some critical $H_{c2}$ where the ground state is
fully ferromagnetic but differs from the XXX ferromagnet in that it
 has a finite gap to the first excited state.
The intermediate region (between $H_{c1}$ and $H_{c2}$) is completely
gapless  and the pitch vector $Q$ (the value of the momenta at which the
static structure factor shows a peak ) decreases continuously
from $ Q=\pi$ at $H_{c1}$ to $Q=0 $ at $H_{c2}$.
Evidence for such transitions were seen in the  spin-$1\over 2$ ladder
compound $Cu_2 (C_5 H_{12}N_2)_2 Cl_4$ \cite{LEVY}
 and the quasi one-dimensional spin-$1\over 2$ systems $TTF-CuBDT$ \cite{TMT}.
The  C-IC transition seen in 
$CuGeO_3$ \cite{CU1,CU2,CU3} can be understood in a similar manner if it
is treated as a spin chain with spin-Peierls interactions alone.  However, 
because of the
presence of strong phonon interactions in $CuGeO_3$, various complications
can arise, in particular, the transition becomes first-order.

In this paper,  we are concerned with the  properties of a generic
gapped spin-$1 \over 2$ system in the presence of an external
magnetic field. Though the naive expectation is that all gapped systems might
exhibit similar behaviours in the presence of the field,
 we find that they  have very different
properties depending on the nature of the interaction which creates the
gap. In particular we focus on experimentally
measurable quantities  like  the total magnetisation, NMR relaxation
rates  and neutron scattering intensities. 
To do this we calculate  various spin-spin  dynamic  correlation
functions  in the gapless region between $H_{c1}$ and $H_{c2}$. We calculate
various exponents in the IC regime close to $H_{c1}$ and use these to calculate 
the temperature dependence of the NMR rates $(T_1)$. We also discuss in detail
the consequences of our results for the dimerised, frustrated and ladder
systems.
We find that though magnetisation measurements will not help differentiate between
these systems, neutron scattering and NMR can. 
Since the models studied in this paper directly describe the compounds mentioned
above, the results obtained here are of immediate relevance to  
experiments. 
 
 The paper is organised as follows:
Sec. II contains a very brief review of
spin-$1 \over 2$ chains and the technique of bosonization used in this paper.
In Sec. III , we introduce our hamiltonian for a generic gapped spin-$1 \over2
$ chain and discuss the effects of the applied magnetic field and
the consequences of incommensurability induced by the magnetic field. We
present in Sec. IV our results for the dynamic correlations and  the values of
the exponents near the transition at $H_{c1}$ for a generic gapped spin-$1
 \over2$ chain. We also discuss the generic phase diagram as a function
of magnetic field and temperature.
The spin ladder is treated separately in Sec. V. In Sec. VI, we
calculate the temperature dependence of the NMR rate $T_{1}^{-1}$ and discuss the
behaviour of the same for the cases of the dimerised, frustrated and
ladder models. Sec. VII contains the concluding remarks. 
\vskip .5 true cm

\section{Introduction to Spin-$1 \over 2$ Systems}

Here we present a brief introduction to the physics of spin-$1\over 2$ 
chains  and summarise certain  results  relevant to the work
presented here.
The main hamiltonian of interest is the Heisenberg model: 
\begin{equation}
{\cal H}= J \sum_{i}\vec S_{i} \cdot \vec S_{i+1} 
\label {1}
\end{equation}
where the $\vec S_i$ is a localised spin $1 \over 2$ operator.
We set $J=1$ in the rest of the calculation. 
This model is Bethe ansatz soluble and is 
known to be gapless. A lot is known about the static and dynamic
properties of this system. Eventhough there are various methods used to
study spin-$1 \over 2$ chains, in this paper we use the machinery of bosonization since this
enables us to deduce various properties in a relatively easy manner.
Since this is an oft used method  we sketch the details briefly.
 We first use the
Jordan-Wigner transformation  \cite{AFF,FDM} which essentially   maps the
spin problem onto a problem of interacting fermions on a lattice. 
For the spin-$1 \over 2$ system considered here, the corresponding
fermionic problem has fermi momentum $k_F = {\pi \over 2}$. We then
perform a
linearisation  around the free fermi points given by
$\pm k_F$ , to obtain  an  effective low energy continuum fermionic theory
and then bosonize using the standard
dictionary of abelian bosonization \cite{AFF,FDM}. We just present
the final expressions obtained for    
  the spin $1 \over 2$ operators in terms of
the bose field $\phi$ and its dual $\theta$ 
\begin{eqnarray}
S^z (x) & =  &
(-1)^x \cos ( 2 \phi ) + {1 \over {2 \pi}} \partial_x \phi
\nonumber \\
S^- (x)& =& \exp (i \theta) [ (-1)^x + \cos (2 \phi )] 
\label{2}
\end{eqnarray}
  Finally, Eqs.(2) can be used to obtain the bosonized version of the hamiltonian given in
Eq.(1) i.e., 
\begin{equation}
{\cal H}= {1 \over  {2\pi}} \int dx [ uK (\pi \Pi)^2 + ({u \over K}) (
\partial_{x} \phi)^2] 
\label {3}
\end{equation}
where $u$  is the spin wave velocity, $\Pi$ is the momentum conjugate
to the field $\phi$  and $K={1 \over 2}$ for the isotropic hamiltonian
of Eq.(1). $\cal H$ is just
the hamiltonian for free bosons. 
Note that $K=1$ for the case of the XX- antiferromagnet which in turn is 
 equivalent to a theory of free fermions via the Jordan-Wigner
transformation. Other values of $K$ 
correspond  to having a $J_z$ coupling and hence to interacting fermions. 
Since there
is no mass term for the $\phi$ field, it is clear that there is no
gap to the first excited state from the singlet ground state. 
Since a free boson theory given by Eq.(3) is trivially solvable, it is fairly
straightforward to calculate the dynamic correlation functions  using Eqs.(2).

\section {Spin-$1 \over 2$ Systems With a Gap in a magnetic field}

\subsection { Zero Magnetic Field Case}

In many compounds which mimic the behaviour of spin-$1 \over 2$ chains,
it has often been found that there are many other interactions between
spins, apart from the the isotropic interaction summarised in the
hamiltonian of Eq.(1). These could be anisotropies, frustration, impurities
and interchain couplings. More often than not, it is found that these interactions are small
and can be 
treated as perturbations  of the above gapless system.
Within the framework presented here, the effects of these
perturbations can be gauged from whether they are relevant, marginal
or irrelevant in the renormalisation group sense.
Depending on whether the perturbations are relevant or irrelevant, a
gap opens in the spectrum. In this paper we are interested in a weak
spin-Peierls dimerisation and  frustration arising from next nearest
neighbour interactions.  

We first consider  a small dimerization $\delta $, which arises from
spin-lattice interactions. The corresponding  perturbation to the
hamiltonian given in Eq.(1) is 
\begin{equation}
{\cal H}_D=  (-1)^i \delta
\vec S_i \cdot \vec S_{i+1}  
\label { 4}
\end{equation}
\noindent
Using Eqs.(2) the  bosonized continuum version 
of (4) is\cite{AFF}
\begin{equation}
{\cal H}_D  = \delta \int dx \cos (2\phi)
\label {5 }
\end{equation}
 It is equally interesting to study the effect of a competing next nearest neighbour  interaction $J_2$. Such an interaction tends to frustrate the system.
The addition to Eq.(1) is
\begin{equation}
 {\cal H}_F= J_2 \vec S_i \cdot \vec S_{i+2} 
\label {6}
\end{equation} 
\noindent
For classical spins,  the ground state retains its
 Neel like order for all $J_2 < 0.25$ and exhibits spiral order  for
$J_2 >0.25$. This is just a crossover as it is not possible for
$J_2$ to drive a zero temperature phase transition in the classical system. What is the
behaviour of the quantum spin-$1 \over 2$ system ? To study this we
again take recourse to Eqs. (2) and write down the bosonized version
of Eq.(6)\cite{AFF} 
\begin{equation}
{\cal H}_F = (J_2 - J_{2c})\int dx \cos (4  \phi)
\label {7}
\end{equation} 

Apart from the cosine terms mentioned above, these  spin interactions  also
 renormalise the  velocity $u$ in (3). 
We now analyse the effects of the interactions given by Eqs.(6,7).
In the free boson theory of
Eq.(3), the anomalous
dimensions of the operators $\cos (n \phi)$ are given by ${n^2 K} \over 4$.
Using this at the isotropic point $K = {1 \over 2}$, we see that
${\cal H}_D$ given by Eq.(5)
has dimension $1\over 2$
and hence  is  relevant for
all values of $\delta$. 
Similarly  ${\cal H}_F$  described in Eq.(7) is found to have dimension $2$
and is marginal.
Numerically  it has been shown that  ${\cal H}_F$ is marginally relevant for
$J_2 > J_{2c}=0.2411$ and irrelevant for $J_2 < 0.2411$ \cite{J2C}.
 Therefore, we can see  that both $\delta$ and
$J_2$ drive a 
quantum phase transition from a gapless phase to a phase where  a gap opens in the dispersion for all
values of $\delta$ and $J_2 > J_{2c}$. 

Instead of restricting ourselves to the dimerised and frustrated models in this paper, we study the following  hamiltonian generic to gapped
spin-$1\over 2$ chains in a magnetic field:
\begin{equation}
{\cal H_{\rm {gap}}}= {1 \over  {2\pi}} \int dx [ uK (\pi \Pi)^2 + ({u \over K}) (
\partial_{x} \phi)^2 +\nu \cos (n \phi) ]
\label {8}
\end{equation}
\noindent
The 2-parameter space spanned by $K~ {\rm and} ~ n$ correspond to various
spin-$1 \over 2$ systems with gaps in their spectra. For example,
 $K= {1 \over 2},n=2$ corresponds to the dimerised model and
$K={1 \over 2},n=4$ is the frustrated antiferromagnet. 

 We  see that
 though all these models have a gap, we can differentiate between them  by 
studying their
excitation spectra  the nature of which depends on the values of $K$ and $n$.
For instance, 
it 
is known from the study of excitations   
in the  sine-Gordon theories  \cite{DHN},  that 
apart from the magnons and solitons there exist certain excitations called
breathers. These are non-linear like the solitons  and  manifest
themselves as discrete levels in excitation spectrum. These breathers have
energies higher than the singlet-triplet gap but lie below  the continuum
part of the spectrum  generated by the solitons. 
The number $N$ of
such breather modes is given by $N < (8 - n^2 K)/{n^2 K}$. For example, we see
that the frustrated model has no breather modes in its excitation spectrum.
On the other hand  the
spin-Peierls system has two breather modes in
its spectrum: the $N=1$ breather is just a renormalized magnon and the
$N=2$ breather is a bound state of two magnons. This extra bound state
is found to affect the spin correlation function at finite temperatures
\cite{MAKI}.
Optical experiments on the lines of Ref.\onlinecite{REI} where breathers
in quasi-1D ferromagnets were detected, can be used
to directly detect these breathers.

\vskip .5 true cm

\subsection{ Effects of a Magnetic Field }

We now  study the effect of a magnetic field on these
gapped systems. To do so,
we  turn on a magnetic field $H$ in the $\hat z$-direction.
Note that the magnetic field breaks the $SU(2)$ symmetry. 
The interaction  with the
spins on the lattice is
\begin{equation}
{\cal H}_m  = \sum_{i} g\mu _{B} H S^z _{i}
\label{9 }
\end{equation}

  A simple understanding of the effect of
the field can be obtained  in the fermionic picture.
 First of all, the application of the magnetic field is
equivalent to a chemical potential for the fermions.
When $H=0$  we have zero uniform magnetisation i.e., $\sum_{i} S^z _i=0$.
This corresponds to a completely filled lower band for fermions
with a gap $\Delta$~~separating the lower and the upper bands and the fermi energy
lying in the middle. As we increase $H$, we are in effect
shifting the fermi energy. A point is reached where the 
fermi energy  crosses the gap and  lies at the bottom of the upper band.
This value of $H$  corresponds to $H_{c1}$ 
i.e.,
$\Delta = g\mu_{B} H_{c{1}}$.
When $H$ is increased beyond $H_{c1}$, the upper band is partially filled
resulting in a non zero magnetisation $m$.  The zero magnetisation
region with a gap below $H_{c1}$ is the commensurate (C) regime  in that, short range
antiferromagnetic order still persists. Above  $H_{c1}$, the ground
state  is magnetised, has no gap  and canted. 
This is what we call the incommensurate (IC) phase. This is a 
 quantum phase transition and $S^z$ is the relevant order parameter.
This is what makes it different from the incommensurate(spiral) phases seen in frustrated
systems (Sec. IIIA) \cite{RC,TH}.

Now that we have a heuristic understanding of the effect of the magnetic
field  we now proceed to calculate various quantities of interest like 
the magnetisation and exponents in the vicinity of $H_{c1}$, within the
continuum approximation.
Using Eqs.(2) we can see that Eq.(9) corresponds to adding a gradient term ~$g \mu_{B} H \partial_{x} \phi$~
to the
hamiltonian of Eq.(8).
\begin{equation}
{\cal H_{\rm {tot}}}= {\cal H}_{gap} + {1 \over  {2\pi}} \int dx 
 g\mu_B H \partial_x \phi
\label {10}
\end{equation}
\noindent
This hamiltonian has the same form as the one used 
in the context of the C-IC transition in two-dimensional systems
where  it is a transition from a phase with a
 a discrete symmetry (C) to a phase with a continuous symmetry(IC)
\cite{POK,HE2}. 
 Note that
the gradient term in Eq.(10) can be eliminated by a simple shift of the $\phi$ field
i.e., $\phi \to \phi + \pi K g \mu_B H x$. Since the only effect of 
$H$ below $H_{c1}$ is to renormalise the gap, we can replace this
shift by $\phi \to \phi + \pi m x$ where $m$ is the magnetisation.
The cosine term, however, is not invariant under this shift. 
From the analogy with the chemical potential in fermionic systems, we
infer that the magnetic field changes the fermi momentum. We can redefine new
fermi points $\pm k^{\prime}_F$, linearise around these points and
obtain a new effective massless free boson theory, albeit with
a different value of $K$. 
 It has gapless modes at $q=0$ and
$q=2k^{\prime}_F=\pi (1 -2m)$.
 A schematic plot of the field dependent dispersion
for the spins in the IC phase is given in Fig.1. 
The behaviour  in the gapless IC region is governed by the hamiltonian
\begin{equation}
{\tilde {\cal H}}={ 1 \over {2\pi}} \int dx [v  {\tilde K} (\pi \Pi)^2 + ({ v \over {\tilde K}}) (
\partial_{x} \phi)^2]
\label{11 }
\end{equation}
\noindent
The quantities $ v$ and $\tilde K$ are dependent on $H$. Although 
it is difficult to obtain  the  dependences  for the entire range of $H$ between
$H_{c1}$ and
$H_{c2}$, one can compute  the exact values
for the exponents (equivalently the $\tilde K$) at the critical point $H_{c1}$  and  see how they
change as we  move away from these points.

Before we proceed with the calculation of $\tilde K$, we  first study 
how the magnetisation rises above $H_{c1}$.
Using the results on the C-IC transition  \cite{HE2}
\begin{eqnarray}
m =& \eta & {\sqrt {K} \over n} \sqrt{ (H^2 - H_{c1}^2)} \nonumber \\
 =& \eta  & {\sqrt {K} \over n} \sqrt{ (H - H_{c1})(H+H_{c1})}\simeq 
\eta {\sqrt {K} \over n} \sqrt{(H- H_{c1})2 \Delta }
\label{12 }
\end{eqnarray}
\noindent
where $n$   is the coefficient of $\phi$ in the argument of the cosine interaction and $\eta$ is a constant of proportionality which depends on the renormalised
velocities.  
The magnetisation increases from its zero value as a square root near $H_{c1}$. 
A similar square root behaviour is seen near $H_{c2}$ if one approaches
this critical point from the ferromagnetic side. Here it is the decrease in
magnetisation from the full ferromagnetic value that shows the square  
root behaviour.  In Fig.2, we have shown the behaviour of the magnetisation
 with temperature and the field in units of $g\mu_B$. This was done using the  analogy with the case of fermions with a gap and a chemical
potential. We assumed a dispersion for the fermions of the form $ \omega_k = \pm \sqrt{(J k)
^2 + ( {\Delta })^2}$. 
Here the  two square root regimes  near $H_{c1}$ and $H_{c2}$ ($0.4$ and $1.1$ in the figure)
 are joined by a region
in which the magnetisation increases in a  nearly linear fashion with $H$.
The interval of $H$ in which  the square root behaviour of the magnetisation
is seen  and also the range where the nearly linear regime is seen, 
depend on the parameters like the couplings and the gap. For certain ranges of
 these parameters where the gap  $\Delta$ is of the order of the exchange 
couplings one obtains
a scenario where the width of the square root regime is
quite small so that even experiments done at reasonably low temperatures could entirely miss the
detection of this region. This might be the reason as to why the
square root behaviour is not observed in Ref. \onlinecite{LEVY}.
Since the behaviour of the magnetisation  given by Eq.(12) is generic to all the gapped
systems studied here, it follows that a measurement of the
magnetisation will not be able to qualitatively differentiate between the
various models.

The magnetisation derived in Eq. (12) applies to the ladder also. This is
because,  as
we will later show in
Sec. V, the effective hamiltonian for the ladder has the same form
as given in Eq.(10). Another system which exhibits a similar transition
in the magnetic field is the spin-$1$ chain.
Since the  spin-$1 \over 2$ ladder  and the spin-$1$ chain belong to the
 same universality class\cite{HE1},  Eq.(12)
describes the magnetisation in the spin-$1$ chain also. 
This agrees well with the  result for $m$ in spin-$1$ chains  obtained 
in Ref.\onlinecite{HE1,IA2}. 
In Ref.\onlinecite{IA2} a mapping of a phenomenological hamiltonian 
of interacting magnons
  onto a system of bosons
with repulsive $\delta$-fn interactions was used to obtain the magnetisation
and the correlations in the IC regime.

Naively one might expect that these gapped systems 
in a magnetic field   
have the same qualitative behaviour for correlation functions and   that 
 quantitative features like $H_{c_1}$ and thermodynamic quantities like 
$m$ are model dependent.
We shall see below that this is not true and that these models have
dynamically different physical behaviours in the IC regime 
depending on the value of $n$. 
For a generic interaction of the form  $\cos (n \phi)$ (where
 $n^2 K \leq 8$ ) and $H ~{\rm close ~to}~ H_{c1}$  on the IC side,
\cite{HE2}
\begin{equation}
{{\tilde K}}= { 4 \over {n^2}}  (1 - {{ u  m \gamma \sinh (2\theta)} \over \Delta}) 
\label{13 }
\end{equation}
\noindent
Here $\Delta$ is the gap and $\gamma$ is a positive constant which depends on the parameters of the
theory but is independent of $H$. $\theta$ is defined by $\exp (-2 \theta)
={{n^2 K} \over 4} $.
At the transition $H_{c1}$, $\tilde K$ goes to a universal
 value $ 4 \over {n^2}$
independent of the value of $K$.  
As $H$ increases,  there is  a non-zero magnetisation $m$ 
and the change in  $\tilde K$  
is completely governed by $n$ and
$K$ in that $\tilde K$ could increase or decrease depending on the
value of $K$. 
For example, 
for the frustrated ($J_2$) model    
$\tilde K= {1 \over 4}$ at $H=H_{c1}$  and increases as $H$  increases  
and for the case of the dimerised $(\delta)$ model, $\tilde K= 1 $  at $H=H_{c1}$ and
decreases with increasing $H$.
Here we have seen that  though the magnetisation has the same qualitative
behaviour for all the models, the value of $\tilde K$ is model dependent
because it is determined by the
 anomalous dimensions of the perturbing operators. 
 As a consequence the exponents in these models
will be radically different as will be shown in the following section.
\vskip.5 true cm

\section{Correlation Functions in the Critical Region}

In this section, we first study how the  structures of the correlation
functions are altered by the magnetisation in the IC phase. Later we
will calculate the various commensurate and incommensurate contributions to
the finite temperature dynamic susceptibility.
From the 
magnetic interaction given  by Eq.(9) and the field shift $\phi\to \phi
+\pi m x$,
 it is clear that               
 since for  $H < H_{c1}$  there is no net magnetisation,
 the form of the  correlators is unaffected and only the gap is
renormalised.  However, in the IC phase, the presence of a non-zero
magnetisation $m$ results in
  $\phi \to \phi + \pi mx$. The dual field $\theta$ is insensitive to
this shift. 
Incorporating this field shift in the expressions for the spin operators given
in Eq.(2), we find that the generic form of the spin-spin correlation
function in the IC region is   
\begin{eqnarray}
\langle S^z (x,t) S^z (0,0) \rangle & = 
 & m^2 + f_1(x,t) + \cos \pi (1-2m) f_2(x,t)
\nonumber \\
\langle S^+ (x,t) S^- (0,0) \rangle & =  &  \cos (2\pi mx) g_1(x,t) + \cos (\pi x)
g_2(x,t)
\label {14}
\end{eqnarray}
\noindent
where $f_1,~f_2,~g_1 ~\rm {and}~ g_2$ are monotonic, decreasing, power-law functions  of
space and time  and they go to zero asymptotically. 
From the above expressions we deduce  the following:
 the correlation function  parallel to the field i.e., 
$\langle S^z S^z \rangle$, has a uniform magnetisation
 or equivalently a  $Q=0$ mode  while the staggered part is shifted from $Q=\pi$ to
$Q= \pi - 2\pi m$.  This is in contrast to the correlation in the plane 
perpendicular to the field, $\langle S^+ S^-\rangle$,
 where the staggered mode is unshifted and
remains at $Q=\pi$  and the  uniform magnetisation ($Q=0$) mode is shifted 
to  $Q= 2 \pi m$. We call $Q=2 \pi m$ and $Q= \pi (1-2m)$ the incommensurate
modes. 
 
With the results obtained above , we now proceed with our calculation of 
experimentally relevant quantities like the dynamic susceptibilities at
finite temperatures.
As a first step we  compute the
associated unequal time correlation functions using Eqs.(2),(14) and
 the result of Eq.(13).
The correlation functions are found to be as follows:
\begin{eqnarray}
\langle \tilde S^z (x,t) \tilde S^z (0) \rangle &=&   
\langle ( S^z (x,t)-m)( S^z (0)-m) \rangle \nonumber \\
&=&   
\cos ( \pi x (1- 2m )) 
(x^2 -t^2)^{-\tilde K} 
 +cste. {\tilde K \over {4 \pi ^2}}( {1 \over { (x- t)^2}} + {1 \over
{(x+t)^2}}) \nonumber \\
 \langle  S^+ (x,t) S^- (0) \rangle &=&  (-1)^x (x^2 -t^2)^{- {1 \over {4\tilde K}}}
+
cste. \cos (2 \pi m x)~~ (x^2 -t^2)^{-({{1 \over {4\tilde K}} +\tilde K -1})}\cdot \nonumber \\
   & & [ \exp (2i \pi mx) 
{ 1 \over {(x-t)^2}}  +
\exp (-2i \pi m x) {1 \over {(x+t)^2}}]     
\label{15}
\end{eqnarray}
\noindent
with $\tilde K$ being specified by Eq.(13).
 It is not surprising that the exponents 
are different in the directions parallel and perpendicular
to the external field. This is because the magnetic field breaks
the $SU(2)$  spin rotational invariance.
For  the case of the dimerised model at the critical  magnetic field 
$H= H_{c1}$, we found in Sec. III that $\tilde K =1$. An interesting coincidence is that
the exponents and hence  the correlation functions calculated here for the 
dimerised model at $H=H_{c1}$
are the same as that for the XX- antiferromagnet  in a zero magnetic
field. The correlation functions of the XX-model (which is in turn
equivalent to a theory of free fermions) can be obtained
by substituting $m=0$  and $\tilde K =1$ in Eqs.(15). Note that this correspondence holds
only at the critical point. 

Using the above results
we can also evaluate the $ (q,\omega)$ dependent susceptibility at finite 
temperatures, both in the direction of the applied magnetic field
and in the direction perpendicular to it. A knowledge of this 
quantity helps us extract various measurable quantities like  
neutron scattering intensities, absorption and nuclear magnetic 
resonance (NMR) rates. The susceptibilities are given by the
following expression
\begin{equation}
\chi_{ij}(q,\omega,T)= -i \int dt~dx~ \exp {i(\omega t - qx)} \theta (t) \langle
[S^i(x,t), S^j(0,0)]\rangle_T 
\label {16}
\end{equation}
\noindent
 Here $i,j$ refer to the components of
the spin and the subscript $T$ implies that the correlator is evaluated
at finite temperature.
Since rotations in the $x-y$ plane still leave the hamiltonian 
invariant, there are
no cross correlations i.e., $\langle S^i S^j \rangle =0 ~\rm {for}~ i \neq j$.
For the susceptibility $\chi_{zz} \equiv \chi_{\parallel}$ in the $\hat z-$direction parallel to the
direction of the applied field \cite{HE1,CF}
\begin{eqnarray}
{\chi}^{Q=0}_{\parallel} (q,\omega,T)&=&  {{q^2} \over {(vq)^2 - \omega ^2}}
\label{17}  \\
{\chi}^{Q=\pi (1-2m)}_{\parallel
}(q,\omega,T)&=& N T^{2\tilde K -2} B( {\tilde K \over 2} -
i {{(\omega+ v(q-Q))} \over {4 \pi T}} , (1-\tilde K))  
B({\tilde K \over 2} -i {{(\omega- v(q-Q)} \over {4 \pi T}} , (1-\tilde K))  
\nonumber
\end{eqnarray}
\noindent
For the perpendicular susceptibility  $\chi_{\perp} \equiv \chi_{+-}$ i.e., in the $x-y$ plane 
\begin{eqnarray}
{\chi}^{Q=2\pi m}_{\perp}(q,\omega,T)&=&  -N^{\prime} T^{2\beta } [ B( {{\beta +2} \over 2} -
i {{(\omega+ v(q-Q))} \over {4 \pi T}} , (-1-\beta))  
B({\beta \over 2} -i {{(\omega- v(q-Q))} \over {4 \pi T}} , (1-\beta))  
\nonumber \\
&+&B({\beta \over 2} -i {{(\omega+ v(q-Q))} \over {4 \pi T}} , ( 1-\beta))  
B({{\beta+2} \over 2} -i {{(\omega- v(q-Q))} \over {4 \pi T}} , (-1-\beta))]  
\label{18}
\end{eqnarray}
\noindent
where $2\beta= 2\tilde K + {1 \over {2 \tilde K}} -2$
\begin{equation}
{\chi}^{Q=\pi}_{\perp}(q,\omega,T)= N T^{2\alpha  -2} B( {\alpha \over 2} -
i {{(\omega+ v(q-Q))} \over {4 \pi T}} , (1-\alpha))  
B({\alpha \over 2} -i {{(\omega- v(q-Q))} \over {4 \pi T}} , (1-\alpha))  
\label {19}
\end{equation}
\noindent
where $\alpha= {1 \over {4 \tilde K}}$ , $B(x,y)$ is the  Beta function and
 $v$ is the  effective magnetisation dependent spin velocity. $N$ and $N^{\prime}$ are velocity and hence field dependent pre-factors.
These expressions for the susceptibilites are valid as long $T < (H -H_{c1})$.
This is because  above this temperature the hamiltonian in Eq.(11) obtained by linearising
the fermi surface  around $k_F^{\prime}$ is  no longer valid.

Though we have explicitly calculated the susceptibilites only in the
IC regime for low temperatures, it is nonetheless interesting to study
the behaviour outside this IC phase. For example, one can study
how the system crosses over to the high temperature classical limit as
a function of the parameters of the theory.
We present a   phase diagram as a function of $T$ and $H$ in Fig.3.
This phase diagram  is true  only for $H < H_{c2}$ because at $H_{c2}$ the system
makes a transition to another gapped phase.
For temperatures smaller than the exchange $J$  one obtains four regimes
as indicated in Fig. 3. All the lines indicate crossovers and are not
 phase transitions.
This phase diagram can be easily understood  within the fermionic picture. 
Note that for the fermionised version of spin chains with a gap the complete
 dispersion for the excitations has the form  $\omega_k = \pm\sqrt {J^2 k^2 
+ { \Delta }
^2}$. These two branches constitute the lower and upper bands with a gap $2\Delta$ separating the two bands.
 With the full dispersion, the magnetisation is given
by a function which depends on two parameters $\Delta \over T$ and $H \over T$. The different regimes to be
discussed below will be characterised by the varied behaviour of this
function.
For all 
 $T < H_{c1}-H$
one gets the gapped phase  where the chemical
potential i.e, $H$  lies in the gap and the temperature is still small enough and does not excite particles to the
upper band. This phase is characterised by an exponential decay of all
correlation functions.
When $H >H_{c1}$ and 
     $T< (H- H_{c1})$, the    
chemical potential lies in the upper band and we get the   
the incommensurate or
Luttinger liquid I  regime as was already discussed in Sec. IIIB
(cf. Fig.3). The dynamical exponent is z=1 here. 
The susceptibilites in Eqs.(17-19) are valid in this Luttinger liquid
regime. 

 If  the chemical potential lies at the bottom of the
 upper band and $T\sim 0$ or equivalently when temperature is such that it 
can excite
particles at the fermi level to the bottom of the upper band, 
one obtains 
the quantum critical
regime\cite{SEN} defined  by $\vert H -H_{c1} \vert < T$. 
In the quantum critical region
the physics is governed by the fixed point\cite{SEN} at $H=H_{c1}$ 
 where $\tilde K$ has a
universal value $ 4 \over {n^2}$.
A simple way of understanding the behaviour in this region is presented
below.
Firstly, the curvature of the bottom of the band becomes very important in
that the effective dispersion seen by the excitations is quadratic.
Here one can expand $\omega_k$ in $\Delta$ 
to obtain $\omega_k^{QC} = \Delta + {{k^2} \over {2 \Delta}}$ in the quantum
critical region.
The dynamical exponent $z=2$ in this region.
A direct consequence of the quadratic dispersion is that  
 $\Delta \over T$ and $H \over T$
are  no longer 
independent scaling variables and only occur in the combination ${H-H_{c1}}
\over T$ and 
the magnetisation $m$ which is just the number of fermions in the upper band
 has the scaling form
$m(T)=  T^{1 \over 2} f({{H-H_{c1}} \over T})$. Analogous scaling functions
exist for the spin-spin correlations also. The scaling functions for the
correlators are functions of $k^2 \over T$, $ \omega \over T$ and
${H-H_{c1}} \over T$. Note that $\Delta \over T$ and $H \over T$
are  no longer 
independent scaling variables and only occur in the combination ${H-H_{c1}}
\over T$.
 The  spins are complicated functions of the fermions and this results in
 the scaling functions being 
difficult to obtain for reasons described in Ref.\onlinecite{SEN}. 
The quantum critical behaviour holds only as long as the quadratic form of the dispersion
assumed is valid or in other words, only as long as  temperature 
is such that the bulk of the excitations is confined to the bottom of the upper
band. 
Similar  phases   were obtained 
in Ref. \onlinecite{SEN} in the context 
 of spin-$1$ chains in a magnetic field,
using a phenomenological theory of magnons with repulsive interactions.
Here starting from  a microscopic description, we find that these phases are 
generic to all gapped chains in a field and we have also obtained the exponents
in the Luttinger liquid regime I.

Now  if temperature is increased further,  there is  a crossover to a
fourth region.   
This regime was  not  obtained in Ref.\onlinecite{SEN}
because  the form $\omega_k^{QC}$ was assumed 
for all values of $k$.  Above a
certain temperature  excitations to higher $k$ states occur and one starts probing the deviation from a quadratic
dispersion and the $k^2$ approximation is no longer valid. As a result
the quantum critical scaling for the magnetisation etc., will no longer be
valid above these temperatures. 
The system then crosses over to a new  region where 
 temperature is large enough such that in addition to the 
excitations which now involve states way above the bottom
of the upper band, there are transitions between the lower and the
upper bands.
This implies  that the gap
becomes irrelevant and the effective dispersion becomes linear in $k$.
 This crossover should occur for temperatures $T$ of the order of
 twice the 
gap $\Delta$ but much lesser
than the exchange couplings such that the interaction
becomes irrelevant  and the resulting
behaviour is that of a Luttinger liquid. We denote this regime as
the Luttinger liquid II.
Classical behaviour sets in
for temperatures  greater than  the exchange couplings. 
The widths of all these regions are dictated by the values of the gaps and
the exchange coupling. For instance, the width of the quantum critical
regime is fixed by the gap alone,
and for magnetic fields away from $H_{c1}$
this width is quite small and not as large as purported to be in Ref. \onlinecite{SEN}.
As a consequence, the scaling arguments  apply only in the vicinity
of the fixed point
at $H_{c1}$  where the width of the quantum critical regime is
appreciable and not for fields far away from it. 
In contrast, the width of the Luttinger liquid
II regime is given by the relative sizes of $J$ and $\Delta$. This width
 can be increased or decreased
by tuning the ratio  $\Delta \over J$.
Applying these arguments to the case of spin-$1$ chains where the gap of
of the order of the
exchange coupling $J$, one finds that for small fields  the classical
limit sets in soon and there are no sharp crossovers between the quantum
critical, Luttinger liquid II and classical regimes.      

In the following paragraphs we discuss the physical significance of
the susceptibilites calculated above. The susceptibilities
 are directly relevant to inelastic
neutron scattering measurements where apart from certain magnetic form
factors, the intensity is proportional
to  the $(q,\omega)$ fourier transform of the full spin-spin correlator  $\langle {\bf S}(x,t)\cdot {\bf S}(0,0)\rangle$.
From Eqs.(17), we can see that the scattering intensity in the IC phase
due to the $S^z$ correlator
, obtained as
a function of $\omega$ for some fixed $q$, should in addition to the peak
generated by the massless excitations near $Q=0$ contains an extra peak 
corresponding to the massless modes at the incommensurate  $Q=\pi (1-2m)$. 
Similarly, for the correlations perpendicular to the field, peaks are seen at
the staggered mode $Q=\pi$ and at $Q=2 \pi m$. These peaks are
divergent  at $T=0$ because of the power law correlations present at these values
of $Q$. At finite temperatures the peak heights are finite and are
determined by the $T$ dependence of Eqs.(17-19).  Such incommensurate
features have been
observed in inelastic neutron scattering data 
from copper benzoate 
which is a $s={1 \over 2}$ system  in a magnetic field\cite{BRO}.
Another probe is electron spin resonance (ESR) which can be used
to directly probe the nature of the continuum of excitations at
$Q=2 \pi m$ \cite{PAL}.

One interesting question is whether the propagating modes corresponding
to the uniform and staggered magnetisation are damped at finite temperature
or in other words, is there any thermal broadening? 
This can be answered  by studying the temperature dependence of the imaginary parts
of the
corresponding susceptibilites. First consider
 the case of the isotropic spin-$1\over 2$ chain in zero magnetic field.
 Here it  is  known that the  uniform magnetisation component of the
$S^z$ correlator diverges as $1 \over {x^2}$. The  corresponding
 uniform  and staggered susceptibilities  at finite temperatures are given by 
\cite{HE1,CF}, Eqs.(17) with $m=0$  and $\tilde K = {1 \over 2}$.
Isotropy results in the same  expressions for the perpendicular susceptibility.
In this context  $v$ is the spinon velocity.
The  dependence of $\chi_{\parallel}^{Q=\pi}$
on temperature tells us that  the
$Q=\pi$ mode is damped at finite $T$. 
The temperature independence of the $\chi_{\parallel}^{Q=0}$   implies that there is no damping of 
the $Q=0$ mode at finite $T$\cite{SU}. 
 This can also be understood from the fact that the exponent
$\beta$ in Eq.(18) is zero at $K= { 1 \over 2}$. 
There is no damping of this mode 
 within
the continuum approximation where the fermionic dispersion was assumed to be
linear. 
Taking into account a
small curvature of the fermion dispersion spectrum does not alter this result
because  the height of the peak which is given by the lifetime of the quasiparticles
 in the fermionic
picture  
becomes infinite as $q\to 0$ even at finite temperatures.  
Another way of understanding this absence of damping of the uniform
mode is by noting  the fact that the total magnetisation in all the directions
commutes with the hamiltonian which in turn implies that there is no damping  of this
 mode in the
lattice spin system.

We now discuss the damping of the various modes in  the IC phase.  The
modes are the uniform magnetisation mode $Q=0$, the staggered mode $Q=\pi$
and the incommensurate modes at $Q=2 \pi m$ and $Q=\pi (1-2m)$. 
From Eqs.(17-19) we see that the dominant contribution to the
$Q=\pi$ and $Q= 2 \pi m$ modes is from the perpendicular susceptibility.
As in the Heisenberg case discussed above, the staggered mode at $Q=\pi$ has a damping factor proportional
to $T^{2 \alpha -2}$.  
Similarly the presence of a non-integral exponent $\beta$ in  $\chi_{\perp}^{Q=2 \pi m}$ 
results  in
 the $Q=2 \pi m$ mode being damped at finite temperatures by a
factor proportional to $T^{2 \beta}$. Similarly the behaviour of
the $Q= \pi (1-2m)$ mode is dictated by
 $\chi_{\parallel}^{Q= \pi (1-2m)}$ in Eq.(17) and has a damping factor
$T^{2 \tilde K -2}$. 
We now consider  the $Q=0$ mode. The dominant contribution to the damping of 
this mode  arises from $\chi_{\parallel}^{Q=0}$.
From Eqs.(17), we find that
the parallel susceptibility does not damp the $Q=0$ modes because
 $\chi_{\parallel}^{Q=0}$ is still independent of temperature. 
Because the system is not isotropic we need to check whether there
is a sub-dominant contribution from the 
perpendicular susceptibility  which damps this mode.
To do this we first study the  
damping for other values of $Q$ between $0$ and $2 \pi m$.
Note that this damping can be studied at the resonance frequencies ~$\omega=v
(Q-q)$~ or away from resonance. 
To obtain the leading temperature dependence of the damping of these 
modes at resonance, we substitute ~$q -2\pi m =\delta q$~ in Eq.(18). The resonance frequency
$\omega$ is fixed at $v \delta q$. The low temperature behaviour of
the imaginary part of the susceptibility is given by 
\begin{equation}
{\beta \over 2} \omega^{\beta -1} T^{\beta +1} + {2 \over \beta} \omega^
{\beta +1} T^{\beta -1}
\label {20 }
\end{equation}
\noindent
This expression implies that there is a thermal broadening of the
$Q=0$ or uniform magnetisation mode i.e., ~$\delta q = -2 \pi m$. 
Nevertheless, using the fact that the total magnetisation in the $\hat z-$ direction
still commutes with the hamiltonian even in the presence of the magnetic field
and that the time evolution of  $\sum_i S^+_i$ involves only $\sum_{i} S^+_i$
we conclude that there is no thermal broadening of the uniform mode $(Q=0)$
mode. If $\omega$ is not at 
the resonance frequency then for low $T$, there is  no obvious
thermal broadening and $Im \chi_{\perp}^{Q=2 \pi m}$ is proportional to
~$[v \delta q (v \delta q -2 \pi m)]^{\beta -1}~$. This expression is not
valid  for $~ \omega -v \delta q < T~$ and fails in the vicinity of 
$\delta q=2 \pi m$. From these calculations we see that though bosonisation describes the physics correctly near $Q=\pi$ and
$Q= 2 \pi m$ for the perpendicular
correlations it  does  not describe modes far away from these two points well. 
This is not surprising in view of the fact that by linearising at
$k_F^{\prime}$ we take into account a lot of spurious states at the
bottom of the band. To summarise, we find that the $Q=0$ mode is
undamped whereas the gapless modes at  $Q=2\pi m, \pi(1-2m)$ and $\pi$ 
are all damped.

\vskip .5 true cm

\section{Spin-$1 \over 2$ Ladder}

Another system which exhibits a behaviour akin to the systems
studied above is the spin ladder. Here we consider a ladder with
two identical and isotropic chains.
Let the spins in chain $1$ be labelled $\vec S_1$ and that in chain $2$,
$\vec S _2$. We bosonize this system in the same manner as above. We refer
the reader to Ref. \cite{HE1,TSV,STR} for details.
We adopt the notations of these references
and introduce the symmetric(triplet)  and antisymmetric (singlet)
 combinations of the fields:
$\phi_{s,a} = {{\phi_1 \pm \phi_2} \over {\sqrt 2}}$ and their
respective duals $\theta_s$ and $\theta_a$.
The  hamiltonian for the ladder is
\begin{equation}
{\cal H} = {\cal H}_a+ {\cal H}_s 
\label{21}
\end{equation}
where
\begin{eqnarray}
{\cal H}_a &= & {1 \over  {2\pi}} \int dx [ Ku (\pi \Pi_a)^2 + ({u\over K }) (
\partial_{x} \phi_a)^2 + g_1 \cos ({\sqrt 8}\phi_a) + 2g_2 \cos ({\sqrt 2}
\theta_a)
] 
\nonumber \\
{\cal H}_s &= & {1 \over  {2\pi}} \int dx [ Ku (\pi \Pi_s)^2 + ({u\over K }) (
\partial_{x} \phi_s)^2 + g_3 \cos ({\sqrt 8} \phi_s) ]
\label{22}
\end{eqnarray}
\noindent
where $K= {1 \over 2}$, $g_1=g_2=g_3 = {{ J_{\perp} \lambda} \over {2 \pi }}$ and $J_\perp
$ is the interchain coupling and $\lambda$ some constant.
All the cosine operators are relevant operators of dimension one.
Eqs.(22) have the same form as the hamiltonian for an isotropic spin-$1$
system  written in terms of two spin-$1 \over 2$ operators\cite{HE1}. 
We note that the spin-$1$ system has fixed values for $g_1,g_2$ and
$g_3$ and has no analog of a tunable parameter like $J_\perp$. Nevertheless
the results to be derived below apply to the spin-$1$ system in a
magnetic field.
From Eqs.(21) and (22) we can infer the existence of    gaps for all non-zero
values of $J_{\perp}$  in the spectra of both the
fields  $\phi_a $ and $ \phi_s$ . The ground state of the ladder
is a spin singlet and there exists a gap to the triplet excited state
characterised by $\phi_s$. Therefore, analogous to the dimerised chains
,we expect the vanishing of the
gap and the onset of a gapless incommensurate phase  for some critical
value of the magnetic field.
Since the magnetic field acts on both chains equally, we can easily see from the
bosonization formulas that  $H$ affects  only the $\phi_s$ field 
\begin{equation}
{\cal H}_s \to{\cal H}_s + \sqrt {2} H \partial_x \phi_s 
\label{23}
\end{equation}
Note that this hamiltonian has the same form as Eq.(10). Therefore,
using the results  obtained in Sec.III, we can immediately see that the $\phi_s$
field becomes massless while in the antisymmetric sector the 
   $\theta_a$
field acquires a non-zero expectation value and the $\phi_a$ still has
a gap. As a result the correlations of the
$\phi_a$ field decay exponentially. 
The value of $\tilde K$ for the $\phi_s$ field at $H=H_{c1}$ is 
$\tilde K = {1 \over 2}$. From Eq.(13),we find that  $\tilde K= {1 \over 2}$ above $H_{c1}$ 
also. This is because $\sinh (2 \theta)=0$ for the ladder. However, one should note
that this is valid only in the region close to $H_{c1}$ and provided that the gaps  
and the magnetic field are small 
compared to the intrachain exchange coupling.
Using the above results  we can compute the spin
correlation functions. 
There are two kinds of correlations: correlations within a chain and between chains.
Here again the magnetic field changes the structure of the 
correlation functions in a manner analogous to that  described in Eqs.(14).
We summarise the results for the various correlators below.
Since we are concerned only with the asymptotic behaviours we present only
the power law contributions to these correlators.
\begin{equation}
\langle S_z^r (x,t) S_z^t(0,0) \rangle = m^2 + {1 \over {8 \pi ^2}} [{1 \over
{(x-t)^2}} + { 1 \over {(x+t)^2}}]
\label{24}
\end{equation}
Here $r,t$ denote the chain labels. 
Unlike in the single chain case, in the ladder  {\it the alternating part at $Q=\pi$ which is  now  shifted to $Q=\pi (1 -2m) $ by the
 magnetic field
decays exponentially}. Similarly
\begin{eqnarray}
 \langle S_1^+(x,t) S_1^-(0,0) \rangle & = &
\langle S_2^+(x,t) S_2^-(0,0) \rangle \nonumber \\
& = & (-1)^x ( {x^2-v^2 t^2})^{-{1 \over {8\tilde K}}} \nonumber \\
 \langle S_1^+(x,t) S_2^-(0,0) \rangle & = & i (-1)^x ({x^2 -v^2 t^2})^{-{1 \over {8\tilde K}}} 
\label{25}
\end{eqnarray}
\noindent
Here again {\it the uniform component  is shifted to $Q= 2\pi m$ and decays 
 exponentially}. The fact that  $\phi_a$ is massive results in an
exponential decay of all 
the incommensurate contributions to the
correlations.  
 This exponential decay of all the incommensurate correlations in the ladder
 is in contrast to the single chain
systems studied in the previous sections.
Also note that except for certain exponentially decaying corrections,
the interchain and intrachain
correlators  have essentially the same  asymptotic behaviours. 
As already mentioned, the spin ladder in the magnetic field  has
 the same exponents as that
of the spin-$1$ chain in a magnetic field \cite {HE1,IA2}. 
The dominant contribution to the perpendicular susceptibility for the ladder
has the same form as that given  in Eq.(19) with $\alpha= { 1 \over {8 \tilde K}}$
 and that to the parallel susceptibility by  $\chi_{\parallel}^{Q=0}$
of Eq.(17). At the critical 
point $H=H_{c1}$, the exponent $\alpha= {1 \over 4}$
for the ladder as well as the dimerised model. However, there is one big difference
between the two models. In the dimerised model, the incommensurate
parts of the $\langle S^z S^z \rangle$ and $\langle S^+ S^- \rangle$  also show  power law behaviours
whereas, in the ladder they decay exponentially. 
This has a serious consequence for neutron scattering intensities.
 This is because for the dimerised system, at $T=0$ the power law divergences of
the incommensurate parts of the dynamic correlations will result in a divergent
peak
at the incommensurate wave vector $Q$. On the other hand the exponential
decay of the incommensurate correlations results in  much smaller peaks at
incommensurate $Q$ 
whose finite height and width are determined by the gaps in the $\phi_a$
field. 
Though away from the critical point the exponents for the two models 
are no longer identical, 
 the discussion presented above
 for the neutron scattering still holds.
\vskip .5 true cm

\section{NMR Relaxation Rates}

With the help of the  susceptibilites derived above, we can easily compute  various
quantities that can be studied by neutron scattering and nuclear
magnetic resonance (NMR).  Here, we focus on NMR and in
particular the the spin lattice
 relaxation time $T_1$. 
 The dominant contribution to $T_1$  comes from
the coupling of the nuclei to the lattice spins. Therefore, it is
a good probe to study the nature of the lattice spin system. 
To obtain the temperature dependence of  $T_1$  we use the following formula
in terms of the local susceptibility  to calculate the same \cite{NMR}.
\begin{equation}
{1 \over {T_1 }}= {\lim}_{\omega \to 0} {{ 2k_B T} \over { \hbar ^2 \omega}}
\int { {dq} \over {2\pi}} F_{ij}(q) \chi_{ij} (q,\omega,T)
\label{26}
\end{equation}
Here the $F_{ij}$ are hyperfine form factors and $\chi_{ij}$ has been defined in Eq.(16). 
In general these form factors
are diagonal in $i,j$ and do not vary much with $q$.  
For a system of  non-interacting spins, $ {(T_1 T)}^{-1}$ is a constant.
For interacting spin systems, the dependence on temperature could be
more complicated because the underlying magnetic order plays a
very important role  in that it changes the effective magnetic field
seen by the nuclei.  Examples are the isotropic Heisenberg model
where, $ {T_1 }^{-1}$ goes to a 
 non-zero value 
 as $T\to 0$  and the spin-Peierls system where $T_1^{-1}$ goes to zero at
$T=0$ because of the 
gap to spin excitations. 

We now use the results obtained in the previous section to calculate the
NMR rates.
We first note that the magnetisation $m$ only shifts the
resonance frequency  and does not change the form of the expressions for
$T_1^{-1}$.
Depending on the kind of NMR done, one can probe specific correlations.
This is especially useful for anisotropic spin systems  and also isotropic
 systems in a magnetic field where the perpendicular
and parallel susceptibilites are different.
For instance, if the NMR was done on the nucleus of the lattice spin, then
the relaxation occurs through a contact interaction and $T_1^{-1}$
 depends on $\chi_{\perp}$ alone. On the other hand if it is done
on other neighbouring nuclei in the compound, the relaxation is through
dipolar interactions  and $T_1^{-1}$ depends on 
$\chi_{\perp}$ and $\chi_{\parallel}$. An amalgam of the two methods
will be useful in isolating the two susceptibilites experimentally.

Substituting the expressions for $\chi_{\perp} ~\rm {and} ~ \chi_{\parallel}
$ derived in Eqs. (17),(18) and
 (19) in Eq.(26), we find that a straightforward power counting yields the 
following leading  low
temperature behaviour for the single chain models:
\begin{equation}
[{1 \over {T_1}}]^{single chain}  =  A_{\perp}
T^{{ 1 \over {2\tilde K}} -1} + A_{\parallel} T^{ 2 \tilde K -1} 
+B_{\parallel} T\\               
\label{27}
\end{equation}
\noindent
where $\tilde K$ is given by Eq.(13). $A_{\perp}$,$A_{\parallel}$ and $B_{\parallel}$ are constants independent of temperature.
The suffixes $\perp$ and $\parallel$ refer to the contributions  from
the perpendicular and parallel susceptibilites. 
 It is easy to see that the staggered susceptibilities dominate
in $T_1^{-1}$. For the ladder model, $T_1^{-1}$ is given by
\begin{equation}
[{1 \over {T_1}}]^{ladder}  =  A_{\perp}
T^{{ 1 \over {4\tilde K}} -1}  
+B_{\parallel} T\\               
\label{28}
\end{equation}
\noindent
The contribution coming from the $A_{\parallel}$ term has not been
explicitly written because it goes to zero exponentially as $T\to 0$.

The temperature dependences of these rates for the dimerised and
frustrated models at the critical point $H_{c1}$  are given below:

\noindent
Dimerised $(\delta)$:
\begin{equation}
{1 \over {T_1  }}= ( A_{\delta \parallel }+B_{\delta \parallel})  T + A_{\delta \perp}T^{- {1 \over 2}}
\label{29}
\end{equation}
Frustrated models $(J_2)$:
\begin{equation}
{1 \over {T_1 }} = A_{f\parallel} T^{-{1 \over 2}} + 
(A_{f\perp}+B_{f\parallel}) T
\label{30}
\end{equation}

In Table I, we present the leading low $T$ contributions to $T_1^{-1}$
 at the transition, for the three models considered in this paper. 
There are other
temperature dependent contributions to the NMR rate, but these go to
zero at $T=0$. As mentioned earlier there are two possible scenarios. One is
that the  nucleus probed  does not correspond to the spins and the
interactions are dipolar. Here  at $H=H_{c1}$, the $T_1^{-1}$
diverges as $T^{-{1 \over 2}}$ for the three models. 
  However, the
  divergent behaviour at low temperature in the frustrated model
arises from the parallel susceptibility whereas in the ladder and
dimerised systems it is the perpendicular susceptibility which leads
to the divergent behaviour. This feature can be used to differentiate
between the models as will be discussed in the following paragraph.
As $H$ is increased, we can see from Eq.(13) that 
$\tilde K$ increases for the  
frustrated model  
 and
 the divergence becomes weaker.
Coincidentally $\tilde K$ decreases for the dimerised system and  the divergence of $T_1^{-1}$
becomes weaker too.
For the ladder, $\tilde K$ does not change with $H$ and the divergence persists  and one
has to go to higher fields to see a deviation from the $T^{-{ 1 \over 2}}$ behaviour.
If for some value of $H$, $\tilde K $ decreases to 
${1 \over 2}$ in the dimerised model and  increases to $1 \over 2$
in the frustrated system
, we see that $T_1^{-1}$ does not diverge and $T_{1}^{-1}$ goes to
a constant $A_{\parallel} +A_{\perp}$ as $T\to 0$. This behaviour occurs
because at some point the magnetic field becomes large enough
such that the interaction which generates the gap becomes unimportant and
should recover the exponents for the chain without the interaction i.e.,
the Heisenberg chain where $T_1^{-1}$  is a constant. For fields greater than this value of $H$, the
exponents vary like those of a Heisenberg chain in a magnetic field. 
The exponent approaches 
  that of the $XX$ chain or free fermions for sufficiently
large fields.

Another way of  differentiating between the three models is if the NMR 
involves  the nuclei of the relevant spins. Here, only the perpendicular
local susceptibilites matter and
 a very interesting
picture  unfolds. At $H=H_{c1}$,
 $T_1^{-1}$ diverges as $T^{-{1 \over 2}}$ for
 the dimerised and ladder models  whereas it approaches zero linearly in $T$
for
the frustrated model. 
 Naively, for gapless systems, we would have expected $T_1^{-1}$ to diverge or
 go to a non-zero constant as $T \to 0$ as it does in the case of the
Heisenberg model \cite{SU}. 
 This $T$ dependence in the frustrated  model is indeed strange because
it is reminiscent of the $T_1^{-1}$ rates for  spin chains with
gaps where $T_1^{-1}$ also goes to zero  as $T \to 0$ but exponentially ! 
 A divergence  will be seen in the frustrated system if $H$ is such that
$\tilde K > {1 \over 2}$. 
For increasing $H$  the $T^{-{1 \over 2}}$ divergence 
survives in the ladder but for  dimerised systems 
  $T_1^{-1}$ becomes less and less divergent as $T\to 0$ and later saturates 
 to a constant $A_{\perp}$ at $T=0$
 for some value of
$H$.  
This is very similar to what was seen in the case of dipolar interactions
discussed above. However, note that the saturation values are different in
both cases.
 We also note that the above discussion rests on the fact that
these fields are smaller than $H_{c2}$ which need not necessarily be
the case.
 As mentioned above, for magnetic fields close to $H_{c2}$  where the ground state is nearly ferromagnetic, 
we expect the system to  approach   the free fermion
limit i.e., $ \tilde K \to 1$. As a result $T_1^{-1}$ is expected to  diverge
as $T^{-{1 \over 2}}$ at $H=H_{c2}$ for all the models irrespective
of their values of $n$.

Eventhough both the ladder and the dimerised systems have the same
divergence at the critical point, one can differentiate between these
two systems by studying the non-divergent contributions to $T_1^{-1}$.
For example, for the frustrated model, there exists a correction  to $T_1^{-1}$
proportional to  $T^{2\tilde K + {1 \over {2\tilde K}} -1}$  
. This exponent changes with
increasing magnetic field and corrections of a similar nature do not
exist in the ladder.
These corrections 
should manifest themselves at not too low temperatures. 
However, inelastic neutron scattering should  be able to differentiate between
them as previously discussed in Sec. V. 
At temperatures large compared to the exchange couplings and the gap, from the
analogy with fermions we expect that
$T_1^{-1} = cste$ for all the models studied in this paper. A similar 
behaviour should be seen in the gapped phase i.e., $H < H_{c1}$ also. 
To summarise, we find that the kind of NMR experiments done can
result in drastically different $T_1^{-1}$ for the three models. 

The results derived here can be checked in 
$CuGeO_3$
and  various other quasi $1$D systems. However, with most
compounds being  $3$D  
the results obtained
in this paper are applicable  only in the temperature interval where the
compound is effectively $1$D
and  that there is no 
 $3$D magnetic ordering.
Such a magnetic ordering in $3$D could also result in divergent NMR
rates.
For instance,  
the  onset of $3$D Neel
order at a certain temperature $T_N$, results in 
 $T_1^{-1}$
 diverging as $(T-T_N)^{-{1 \over 2}}$ for $T >T_N$\cite{MOR}. This behaviour
is valid for a temperature range of size $T_N$.
Since this divergence is
the same as seen in the ladder,dimerised and frustrated systems at
$H=H_{c1}$ it is important to establish whether such a divergence arises
from the quasi-one dimensional or the $3$D nature of the compound.
This can be checked by working in the appropriate temperature interval where
there is no $3$D ordering
or by increasing the magnetic field. If the compound
is in the $1$D regime, its exponents vary with the field as predicted
above and it if has $3$D order the exponents do not vary with the
field. 
 
\vskip .5 true cm

\section{Conclusions}

We have  studied the behaviours of various spin-$1\over 2$ models in the 
gapless IC phase induced by an external magnetic field. For a generic
gapped spin-$1 \over 2$ in a magnetic field, it was shown that
the magnetisation is zero below $H_{c1}$ and rises as a square root
above it. 
We found that the gapless behaviour in the IC regime is determined by the dimension
of  the cosine operator and hence different systems have drastically different properties. 
The results presented here were 
obtained from a microscopic theory and not from a phenomenological theory 
as was done in the case of the spin-$1$ chain \cite {SEN}. 
We then discussed the implications of the finite magnetisation for the correlation
functions. 
We found that the effect of the finite magnetisation was to shift the 
$Q=\pi$ mode in  the $S^z S^z$ correlators to $Q=2 \pi m$  and the $Q=0$ mode
 in the $S^+S^-$ correlators
to $Q=\pi (1 -2m)$.
 We  also calculated the unequal time correlation functions  and have provided
explicit formulas for the various susceptibilites as a function of
$T, \omega,q ~{\rm {and~ the~ magnetisation}}~ m$. 
These were  used to study the
thermal broadening of the various modes in a single chain. We find that
 the modes at $Q=\pi$, $Q= 2\pi m$ and $Q=\pi (1-2m)$  are
broadened at finite temperatures whereas the $Q=0$ mode is not. 
Using the susceptibilites we also showed that 
neutron scattering intensities had extra peaks arising from the incommensurability in
single chains but not in the ladder systems. 
We have also calculated the NMR relaxation rates as functions of temperature and have
discussed the results for the dimerised,frustrated and ladder systems in detail. 
Using the fermion analogy, we find  that the
 phase diagram for a generic gapped chain as a function of field and
and temperature has five regions as shown in Fig.3. In contrast to
Ref.\onlinecite{SEN} where the system stays in the quantum critical
regime for a  wide range temperatures before it crosses over to the
classical high temperature limit, we find that the system crosses over
from the quantum critical regime to a second Luttinger liquid regime before 
it becomes classical.  This intermediate temperature behaviours follow
from the form of the dispersion spectrum of fermions with a gap due to
interactions.
Lastly, eventhough
we have given the exact values of the exponents at the critical 
point $H_{c1}$  alone, a knowledge of the magnetisation $m$ from experiments
can be used in conjunction with
 Eq.(13) to obtain the exponents and hence $T_1^{-1}$
close to the transition at $H_{c1}$. 
We find that though magnetisation measurements
cannot distinguish between the various models, techniques like NMR or neutron
scattering which probe the dynamical spin-spin correlations will be able to do so.
Other frequently used methods to study spin systems are EPR (electron
paramagnetic  resonance) \cite{ESR1,ESR2} and Raman scattering. 
 These methods might also be
able to directly differentiate between the dimerised and ladder systems.
We conclude by observing that it should be possible to verify the
results obtained here in NMR measurements being done on $CuGeO_3$ and
 $Cu_2 (C_5 H_{12} N_2)_2 Cl_4$\cite{CHAB}.

\section {Acknowledgements}
We  would like to thank J.P. Boucher, G. Chaboussant and L. Levy
   for interesting discussions
and for  critical readings of the manuscript.
\vfill
\eject

\newpage
\centerline {\bf Table I}\hfill
\centerline{ \bf Temperature dependence of $T_1^{-1}$ at $H=H_{c1}$}
\noindent
\begin{center}
\begin{tabular}{|l|l|l|r|}   \hline

Model       & $\tilde K$ at $H_{c1}$ &~~~ ${T_{1 \parallel} ^{-1}}$~~~  &~~~${T_{1 \perp}^{-1 }}$~~~\\ \hline

Spin-Peierls ($\delta$) & ~~~   ${1 }$ &~~~ $ T$~~~  &~~~$T^{-{1 \over 2}}$~~~ \\

Frustration  ($J_2$)    & ~~~   $1\over 4$ &~~~  $T^{-{1 \over 2}}$~~~ &$T$~~~\\

Ladder                  & ~~~   ${1 \over 2}$ &~~~ $T$~~~ &~~~$T^{-{1 \over 2}}$~~~\\ 
\hline
\end{tabular}
\end{center}
\newpage
\centerline{\bf Figure Captions}  
\begin{itemize}
\begin{enumerate}
\item {Schematic picture of the field dependent dispersion in the IC
phase as seen by $S^z S^z$ correlations (bold line)  and $S^+ S^-$ correlations
(dashed line). $Q=0$ and $Q=\pi$ are the usual commensurate modes 
and $Q=2 \pi m$ and $Q=\pi (1-2m)$ are the
incommensurate modes.}
\item{Plot of the magnetisation $m$ (apart from an overall normalisation)
 versus magnetic field ${H }$ for $\Delta=0.4J$ at various temperatures
$T $. $H$ and $T$ have been normalised by the exchange coupling $J$.
 The solid line 
corresponds to
 $T=0$, the dashed line at $T=0.05$ and the dashed dot line to $T=0.25$.
We see that even at sufficiently low temperatures the square root regime
 gets wiped out 
and $m$ increases in a nearly linear fashion. }
\item {  Phase diagram of gapped spin-$1 \over 2$ chains in a magnetic field $H$
as a function of temperature $T$. There are essentially 5 regions and all the
 lines indicate crossovers between these regions.}
\end{enumerate}
\end{itemize}
\eject
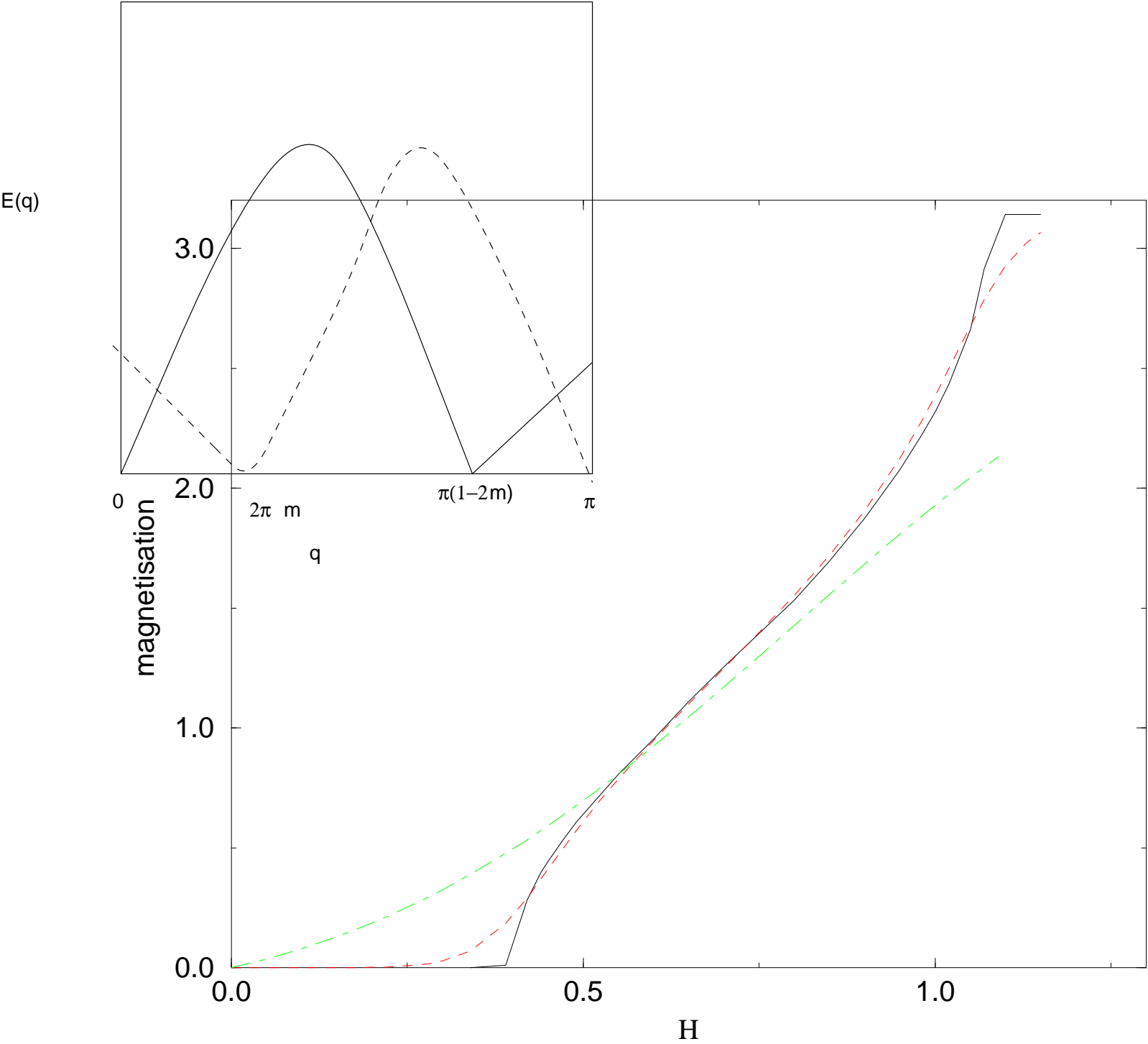
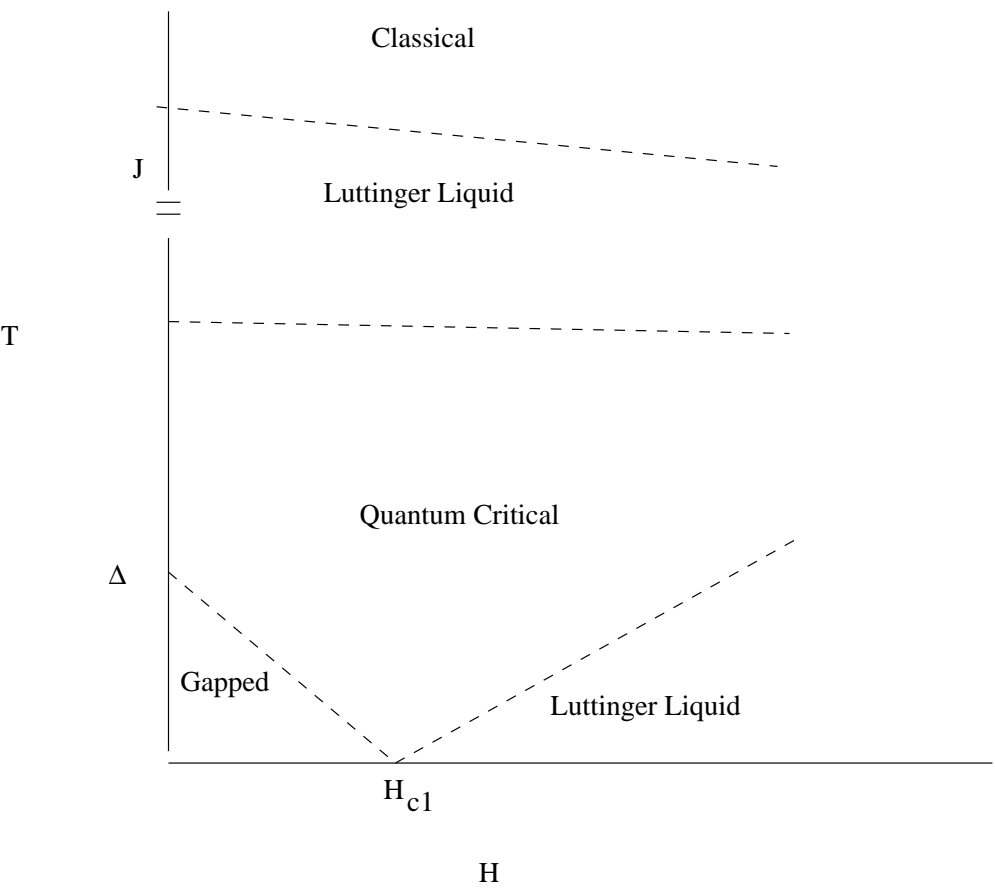
\end{document}